\documentclass[prl,showpacs,twocolumn,aps,superscriptaddress,a4paper]{revtex4}
\usepackage{dcolumn}
\usepackage{amsmath}
\usepackage{graphicx}
\usepackage{latexsym}  
\usepackage{amsfonts}  
\usepackage{amssymb}  
\usepackage{color}  
\DeclareGraphicsExtensions{.pdf,.gif,.jpg,.png}  
  
\newcommand{\be}{\begin{equation}}
\newcommand{\ee}{\end{equation}}
\newcommand{\bea}{\begin{eqnarray}}
\newcommand{\eea}{\end{eqnarray}}
\newcommand{\w}{\omega}
\newcommand{\g}{\gamma}
\newcommand{\tr}[1] { {\rm Tr} \left\{ #1 \right\} }

\begin{document}

\title{Towards a description of the Kondo effect using time-dependent density functional theory} 

\author{G. Stefanucci}
\affiliation{Dipartimento di Fisica, Universit\`{a} di Roma Tor Vergata,
Via della Ricerca Scientifica 1, 00133 Rome, Italy}
\affiliation{INFN, Laboratori Nazionali di Frascati, Via E. Fermi 40, 00044 Frascati, Italy}
\affiliation{European Theoretical Spectroscopy Facility (ETSF)}
 
\author{S. Kurth}  
\affiliation{Nano-Bio Spectroscopy Group, 
Dpto. de F\'{i}sica de Materiales, 
Universidad del Pa\'{i}s Vasco UPV/EHU, Centro F\'{i}sica de Materiales 
CSIC-UPV/EHU, Av. Tolosa 72, E-20018 San Sebasti\'{a}n, Spain} 
\affiliation{IKERBASQUE, Basque Foundation for Science, E-48011 Bilbao, Spain}
\affiliation{European Theoretical Spectroscopy Facility (ETSF)}

\date{\today}  
  
\begin{abstract}
We demonstrate that the zero-temperature conductance of the Anderson 
model can be calculated within the Landauer formalism 
combined with static density functional theory (DFT). 
The proposed approximate functional is based on finite-temperature 
DFT and yields the exact Kohn-Sham potential at the 
particle-hole symmetric point. Furthermore, in the limit of zero temperature 
it correctly exhibits a derivative discontinuity which is shown to be 
essential to reproduce the conductance plateau.
On the other hand, at the Kondo temperature the exact Kohn-Sham 
conductance overestimates the real one by an order of magnitude. 
To understand the failure of DFT we resort to its time-dependent version and 
conclude that the suppression of the Kondo resonance 
must be attibuted to 
dynamical exchange-correlation corrections.
\end{abstract}


\maketitle

Despite the many successes of density functional theory (DFT) 
\cite{FiolhaisNogueiraMarques:03-2} in the description of the electronic structure of 
many-electron systems, the treatment of strongly correlated systems 
is a notoriously difficult challenge. 
However, the fundamental theorems of static and also 
time-dependent (TD) DFT \cite{MarquesUllrichNogueiraRubioBurkeGross:06} are exact. 
Therefore in principle these systems should be accessible and the failure to describe 
them within (TD)DFT has to be accredited to shortcomings of the approximations for the 
exchange-correlation (xc) functional used in practice.

An exact property of the static xc functional is the derivative 
discontinuity at integer occupancy \cite{PerdewParrLevyBalduz:82}.
Recently the crucial importance of this property
in the description of strong correlations has been recognized in, e.g., 
the Hubbard 
model \cite{GunnarssonSchoenhammer:86,LimaOliveiraCapelle:02},
gaps in strongly correlated 
molecules \cite{MoriSanchezCohenYang:09},  Coulomb 
blockade (CB) \cite{KurthStefanucciKhosraviVerdozziGross:10} and the Mott 
transition \cite{kpv.2011}.

For non-equilibrium situations, the development of TD xc functionals 
is still at its infancy. 
Typically one uses the so-called adiabatic approximation by importing 
functionals constructed for static DFT to the time domain.
In the context of quantum 
transport, one of the most debated questions is
whether (or under which circumstances) 
static DFT within the Non-Equilibrium Green's Function (NEGF)
approach can yield accurate conductances.
In a seminal paper, Schmitteckert and Evers 
\cite{SchmitteckertEvers:08} provided strong numerical
evidence that the Kohn-Sham (KS) conductance of single channel 
correlated 
model junctions is in good agreement with the 
exact one. Later, 
the accuracy of the KS conductance has been 
explained and assessed using the Friedel sum rule 
\cite{MeraKaasbjergNiquetStefanucci:10}.
In general, however, the DFT+NEGF formalism is incomplete. 
A proper TDDFT treatment \cite{StefanucciAlmbladh:04-2,ewk.2004} leads to dynamical xc 
corrections \cite{KoentoppBurkeEvers:06}, 
the quantitative importance of which has yet to be clarified.

In this Letter we show that the derivative discontinuity is 
the {\em necessary} and {\em sufficient} ingredient to describe the Kondo effect \cite{Kondo:64}
in the zero temperature conductance using DFT+NEGF.
We propose an analytic KS potential for 
the Anderson model based on finite-temperature DFT \cite{Mermin:65} for an 
isolated impurity. For this approximate potential, the derivative 
discontinuity and, as a consequence also the  conductance plateau, 
emerge naturally in the zero-temperature limit. 
At finite temperatures, instead, we demonstrate that the DFT+NEGF approach is 
{\em not} sufficient. Although our potential is exact at the 
particle-hole (ph) symmetric point, 
at the Kondo temperature the KS and the 
exact conductances differ by almost one order of magnitude. 
We thus give a first explicit example where the exact dynamical xc 
corrections can be quantified and show that they can be
as large as the quantum of conductance.


We consider the Anderson model for a single non-magnetic impurity 
attached to two non-interacting leads. The Hamiltonian 
of this system is given by 
\be
\hat{H}= \hat{H}_{I}+\sum_{\alpha=L,R}\hat{H}_{\alpha}+\hat{H}_{T} \;.
\label{hamil}
\ee
Here, $\hat{H}_{\alpha}=-\sum_{\sigma}\sum_{i=1}^{\infty} (V \hat{c}_{i+1 \alpha, 
\sigma}^{\dagger} \hat{c}_{i \alpha, \sigma} + {\rm h.c.})$ describes, in 
standard notation, the 
tight-binding lead $\alpha=L,R$ 
while  $\hat{H}_{T}=-\sum_{\alpha,\sigma} (V_{\rm link} \;
\hat{c}_{1 \alpha, \sigma}^{\dagger} \hat{d}_{\sigma} + {\rm h.c.})$ 
accounts for the (symmetric) coupling between the impurity and the leads.
In this work we focus on the half-filled system
and take $V_{\rm link}\ll V$. In this parameter range the only 
relevant energy scale for electron tunneling is 
$\gamma=2V^{2}_{\rm link}/V$ (wide-band limit). 
The impurity Hamiltonian is
\be
\hat{H}_{I} = v_0
\hat{n} + U \hat{n}_{\uparrow} 
\hat{n}_{\downarrow}
\label{hamil_dot}
\ee
where $v_{0}$ is the on-site energy (or gate voltage), $U$ is the 
charging energy, $\hat{n}_{\sigma}=\hat{d}_{\sigma}^{\dagger}\hat{d}_{\sigma}$ is 
the number operator for electrons of spin $\sigma$ at the impurity 
and $\hat{n}=\hat{n}_{\uparrow}+\hat{n}_{\downarrow}$. 

\begin{figure}[t]
\includegraphics[width=0.45\textwidth]{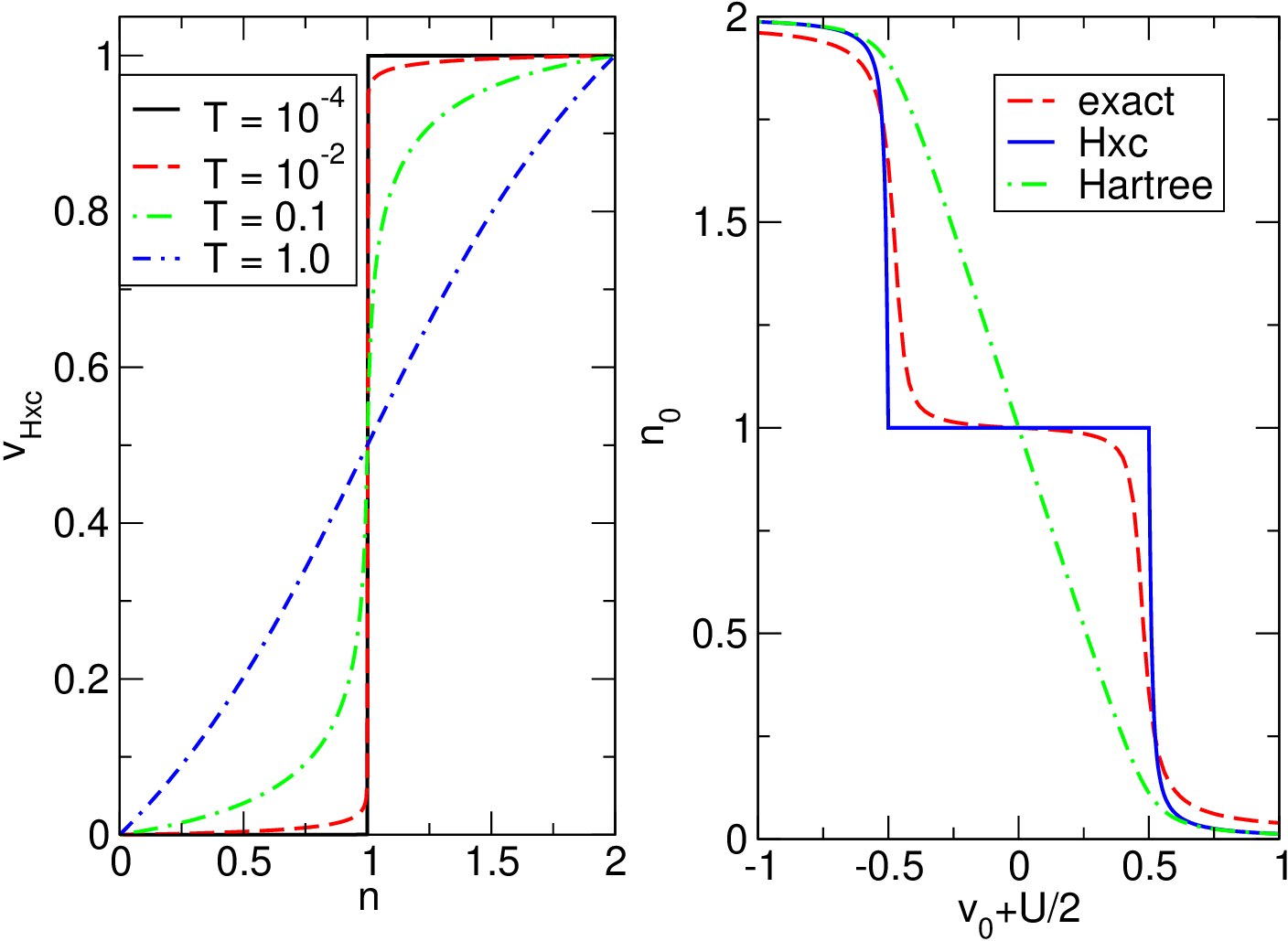}
\caption{Left panel: Hxc potential of Eq.~(\ref{hxc_pot}) 
for different temperatures $T=1/\beta$. Right panel: Hartree and Hxc 
self-consistent impurity density in 
comparison to exact results for $T=0$ and $\gamma=10^{-2}$.
Energies are given in units of $U$.}
\label{vhxc}
\end{figure}

For a DFT treatment of the problem, the first task is to construct an 
approximation to the xc potential. 
Here we propose a KS potential based on an isolated impurity
with Hamiltonian $\hat{H}_{I}$ in contact with a thermal 
bath at inverse temperature $\beta$ and chemical potential $\mu$ \cite{EversSchmitteckert}. 
The non-interacting KS Hamiltonian then reads
\be
\hat{H}_{I}^{\rm s} =  v_{\rm s}  
\hat{n} \; . 
\label{kshamil_dot}
\ee
For both Hamiltonians $\hat{H}_{I}$ and $\hat{H}_{I}^{\rm s}$, the eigenstates 
for electron occupation zero, one, and two are, respectively, 
$|0\rangle$, $|\uparrow \rangle$, $|\downarrow \rangle$, and 
$|\uparrow \downarrow \rangle$ with
eigenvalues  $0$, $v_0$, $v_0$, and 
$2 v_0+U$ for $\hat{H}_{I}$ and $0$, $v_{\rm s}$, $v_{\rm s}$, 
and $2 v_{\rm s}$ for 
$\hat{H}_{I}^{\rm s}$. The density of the 
interacting impurity is
\be
n  = \tr{e^{-\beta (\hat{H}_{I}-\mu\hat{n})}\hat{n}}/Z
\label{dens_pot_int}
\ee
where $Z=\tr{e^{-\beta (\hat{H}_{I}-\mu\hat{n})}}$ is the grand-canonical 
partition function. Equation (\ref{dens_pot_int})
depends only on  $\tilde{v}_{0}=v_{0}-\mu$
and the function $n(\tilde{v}_{0})$ can be inverted explicitly as 
\be
\tilde{v}_{0}(n) = -U-\frac{1}{\beta}
\ln\left(\frac{\delta n +\sqrt{\delta n^{2} +e^{-\beta U}(1-\delta 
n^{2})}}{1-\delta n}\right),
\label{pot_dens_int}
\ee
with $\delta n=n-1$.
The Hartree-exchange-correlation (Hxc) part of the KS potential can then be obtained 
from
\be
v_{\rm Hxc}(n) = \tilde{v}_{\rm s}(n) - \tilde{v}_{0}(n) = 
\frac{U}{2} +g(n-1)
\label{hxc_pot}
\ee
where 
$g(x) = \frac{U}{2} + \frac{1}{\beta} 
\ln \left( \frac{x + \sqrt{x^2 + e^{-\beta U}(1-x^2)}}{1+x}
\right)$.
This is an odd function of its argument, $g(-x)=-g(x)$, and therefore 
$v_{\rm Hxc}(n=1)=\frac{U}{2}$ for all temperatures. We note in 
passing that Eq. (\ref{hxc_pot}) can also be obtained by 
differentiation of the 
Hxc part of the grand canonical potential with respect to $n$. 
The Hxc potential (\ref{hxc_pot}) is shown 
in the left panel of Fig.~\ref{vhxc} for different values of the 
temperature $T=1/\beta$. In the limit $T\to 0$ it becomes a 
simple step function with a step of height $U$ at  $n=1$. Thus, the  
$T=0$ discontinuity of the xc potential emerges naturally from our 
grand canonical DFT treatment.

For 
the Anderson Hamiltonian (\ref{hamil}) we make a local approximation and 
assume that the KS potential  vanishes  in the 
leads while on the impurity is given by $v_{\rm s}(n) = v_0 + v_{\rm Hxc}(n)$. 
We expect this approximation to be accurate for $U/\gamma \gg 1$.
The KS 
Hamiltonian then reads
$
\hat{H}^{\rm s} = \hat{H}_{I}^{\rm s} + \hat{H}_{L} + \hat{H}_{R} + 
\hat{H}_{T}$
with $\hat{H}_{I}^{\rm s}$ from Eq. (\ref{kshamil_dot}). 
Using standard embedding techniques the solution of 
the KS equations are reduced to the self-consistent solution of a 
single equation for the density $n=n_0$
\be
n_0 = \frac{2}{\pi} \int_{-\infty}^{\infty} {\rm d} \omega \; 
\frac{\gamma}{(\omega - v_{\rm s}(n_0))^2 + \gamma^2} 
f_{\beta}(\omega) 
\label{dens_self}
\ee
where $f_{\beta}(\omega)=1/(1+e^{\beta (\omega-\mu)})$ is the Fermi 
distribution. For low temperatures, the Hxc potential  
varies very rapidly for $n_{0}\sim 1$, i.e., in the regime 
in which we are interested (see below). Therefore, instead of solving for  
$n_0$ it is numerically much more advantageous to express 
$n_{0}$ in the l.h.s. of 
Eq.~(\ref{dens_self}) in terms of $v_{\rm s}$ and solve for $v_{\rm s}$.

The resulting self-consistent solution of the density as a function 
of $v_{0}$ exhibits a plateau between $-U$ and $0$ with 
value $n_{0}=1$ (see right panel of Fig.~\ref{vhxc}). This is 
in agreement with the exact solution of the Anderson model 
\cite{WiegmannTsvelick:83}. Using the DFT+NEGF formalism we then 
proceed to calculate the conductance $G$ from
\be
\frac{G}{G_0} = - \int_{-\infty}^{\infty} {\rm d} \omega \; {\cal T}(\omega) 
\frac{\partial f_{\beta}(\omega)}{\partial \omega} 
\label{conduct_zero}
\ee
where 
\be
{\cal T}(\omega)=\frac{\gamma^{2}}{(\w-v_{\rm s}(n_{0}))^{2}+\gamma^{2}} 
\label{kstrans}
\ee
is the zero-bias KS transmission function and 
$G_0=1/\pi$ is the quantum of conductance. 
\begin{figure}[t]
\includegraphics[width=0.47\textwidth]{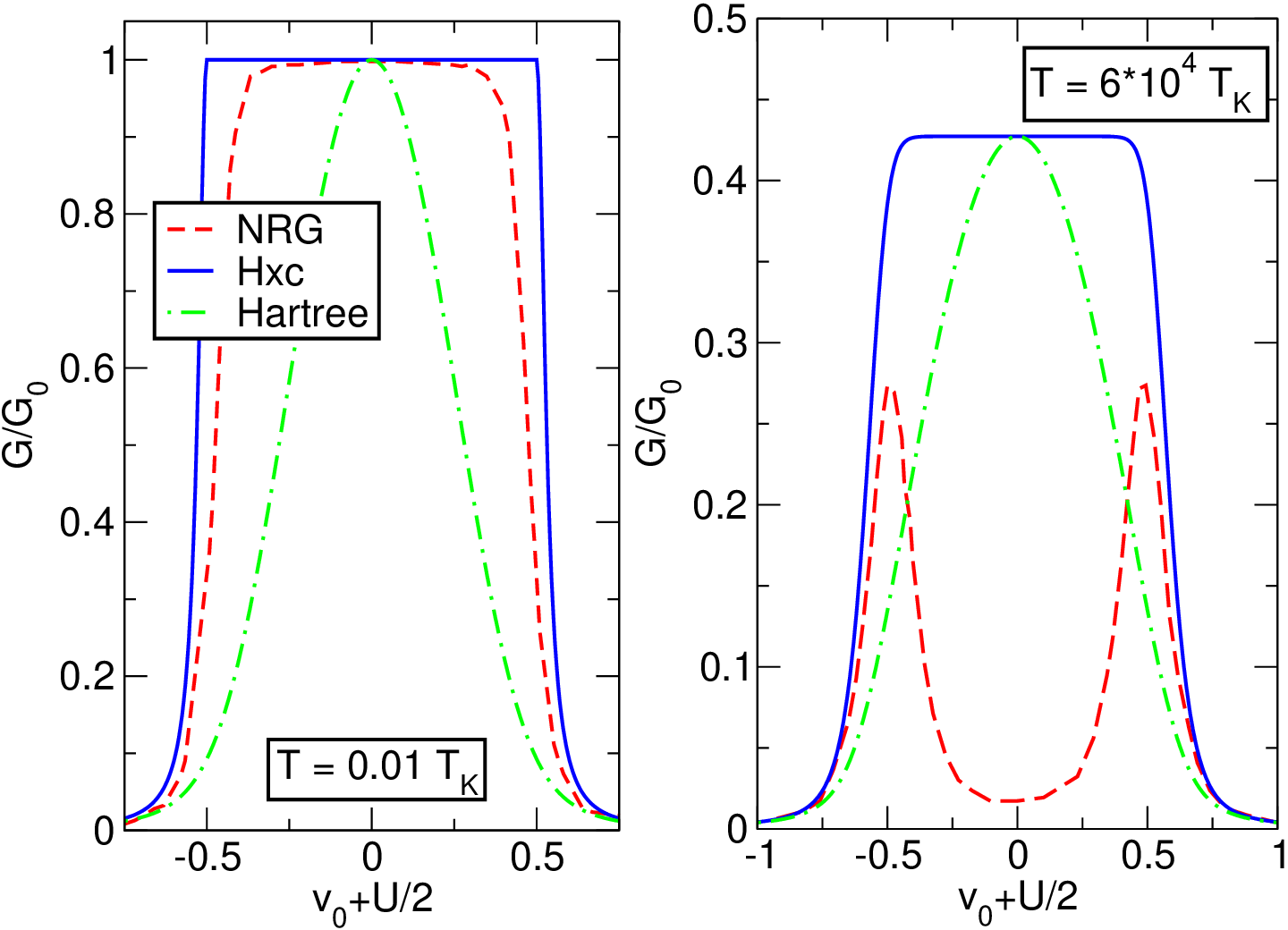}
\caption{Hartree, Hxc and NRG
\cite{IzumidaSakai:05}
conductances versus $v_0$ for two different 
temperatures. The parameters are: $\mu=0$, $\gamma=\pi\times 10^{-2}$
[$T_{\rm K}=6.6\times10^{-7}$]. All energies are given in units of $U$. }
\label{cond_kspot}
\end{figure}

In the left panel of Fig.~\ref{cond_kspot} we plot $G$ as a 
function of $v_{0}$
using the Hxc potential of Eq.~(\ref{hxc_pot})
for $T=10^{-2}T_{\rm K}$, with the 
Kondo temperature
$T_{\rm K} = \sqrt{2U \gamma} \exp\left(\frac{\pi v_0 (v_0+U)}{4 \gamma U}
\right)$ \cite{AleinerBrouwerGlazman:02}. For comparison we also report the 
results within the Hartree approximation, i.e., for 
$v_{\rm s}= v_0 + U n_0/2 $, and the accurate Numerical 
Renormalization Group (NRG) results from 
Ref. \onlinecite{IzumidaSakai:05}. The first remarkable feature is the 
plateau in the Hxc conductance with value $G_0$ in the range 
$-U \leq v_0 \leq 0$, in  good agreement with the exact result. 
This is a direct consequence of the discontinuity of the proposed 
$v_{\rm Hxc}$ at $n=1$ which guarantees that $v_{\rm s}$ is pinned 
to the Fermi energy. In contrast, any
continuous potential, like the Hartree potential, cannot capture the plateau.
Physically, the plateau  cannot be explained solely in terms of 
CB, 
according to which $G$ should be peaked at 
the end-points of the plateau and should instead be small at the ph 
symmetric point $v_{0}=-U/2$.
The value $G=G_{0}$ at $v_{0}=-U/2$ 
is due to the formation of a 
resonant many-body singlet state between the spin of the impurity electron and 
the spin of the screening  cloud, i.e., the Kondo effect. 
The first important finding of this Letter is therefore that
the Kondo effect in $G(T=0)$
is within reach of the DFT+NEGF approach provided that the 
approximate $v_{\rm Hxc}$ has the proper discontinuity at integer $n$. 

The theoretical possibility of describing the  conductance in the 
Kondo regime
within  DFT+NEGF  can be understood  in two  ways. (1) For proportional 
coupling the Meir-Wingreen formula \cite{MeirWingreen:92} 
at $T=0$ yields
\be
\frac{G}{G_{0}}=\g^{2}|{\cal G}(\mu)|^{2}\frac{\gamma-{\rm 
Im[\Sigma(\mu)]}}{\gamma}
\label{mwcond}
\ee
where ${\cal G}^{-1}(\w)=(\w-v_{0}-\Sigma(\w)+i\gamma)$ is 
the impurity Green's function and $\Sigma$ is the many-body 
self-energy. Since quasi-particles at the Fermi energy have an 
infinitely long life-time then ${\rm Im}[\Sigma(\mu)]=0$. From 
Eq. (\ref{mwcond}) we thus see that it is possible to reproduce the 
exact conductance  in a KS system with $v_{\rm 
s}=v_{0}+{\rm 
Re}[\Sigma(\mu)]$. (2) From the Friedel 
sum rule we know that  at $T=0$ the conductance of the Anderson model
is completely determined by the density at the impurity 
\cite{MeraKaasbjergNiquetStefanucci:10}, i.e., $G=G(n_{0})$. Since 
exact DFT yields the exact density
then it must also yield the exact 
conductance. Note that the two explanations above are complementary but not 
equivalent since the exact KS potential is not  rigorously zero 
in the neighborhood of the impurity. It is also worth stressing that 
the equality between the exact and KS conductances does not follow 
from an equality  between the corresponding spectral functions. The 
latter are completely different, although the Kondo peak (of width 
$T_{\rm K}$) and the KS peak (of width $\gamma$) 
both occur at the Fermi energy.

\begin{figure}[t]
\includegraphics[width=0.47\textwidth]{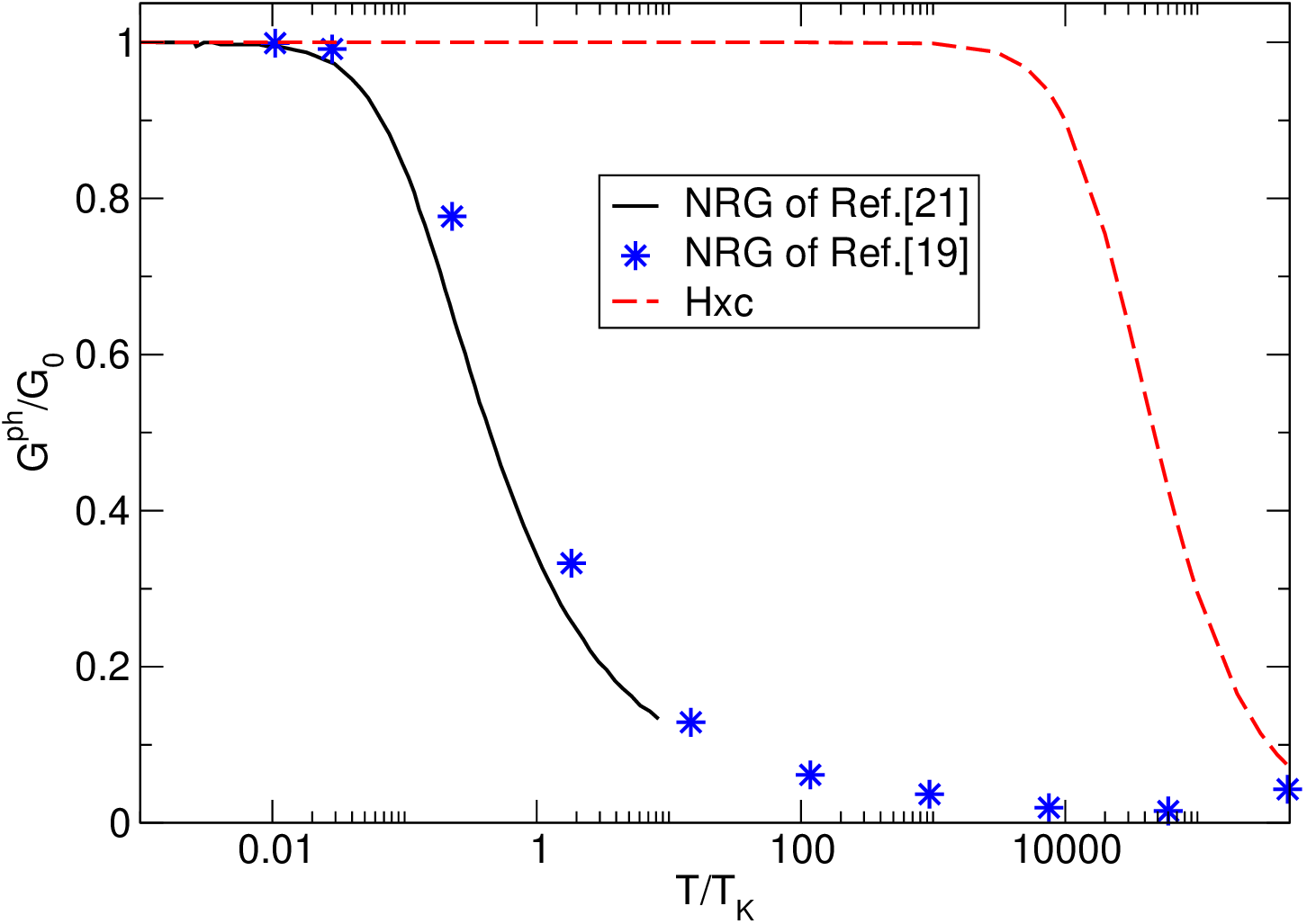}
\caption{NRG conductances from Ref. 
\cite{IzumidaSakai:05} [stars(blue)] and Ref. 
\cite{Costi:00} [solid(black)] against the exact Hxc 
conductance [dashed(red)] at the ph symmetric point versus
temperature.}
\label{cond_ph}
\end{figure}

At finite temperature $G$ does not depend 
on $n_{0}$ alone. 
In the right panel of Fig. \ref{cond_kspot} we display the NRG, Hxc 
and Hartree conductances 
as a function of $v_{0}$ for $T=6\cdot 10^{4} \;T_{\rm K}$.
The failure of the 
Hxc and Hartree approximation in reproducing both the dip at $v_{0}=-U/2$ 
as well as the CB side peaks
is evident.
In Fig. \ref{cond_ph} we compare the Hxc conductance 
with the exact conductance \cite{IzumidaSakai:05,Costi:00} for different 
temperatures at the 
ph symmetric point. While the Kondo 
peak is strongly suppressed for $T\gtrsim 10\; T_{\rm K}$,
the DFT+NEGF approach predicts a conductance $G_{0}$ up to much higher 
temperatures. For $T\gtrsim U$, thermal fluctuations destroy the CB 
and the KS and exact conductances merge and approach  zero for 
$T\to\infty$. 

Is it possible to reproduce the temperature behavior of the exact $G$ by 
improving the quality of the KS potential? The answer to this question  
 is negative since at the ph symmetric point our 
Hxc potential is {\em exact}. For $v_{0}=-U/2$ the Hamiltonian in 
Eq. (\ref{hamil}) is 
invariant under the transformation $\hat{c}_{i\sigma,\alpha}\to 
(-1)^{i+1}\hat{c}^{\dag}_{i\alpha,\sigma}$ and 
$\hat{d}_{\sigma}\to\hat{d}^{\dag}_{\sigma}$, and hence the density is 
unity  
for all sites at {\em all} temperatures. The KS potential which reproduces 
this density is zero everywhere (in leads and impurity) for all $T$ and 
$V_{\rm link}$. 

If one cannot reproduce the exact $G$ by using the exact KS 
potential we must conclude that the DFT+NEGF approach is not sufficient. To trace back the origin of the problem 
we resort to TDDFT \cite{RungeGross:84} and its lattice 
version \cite{combination}. 
In TDDFT it is possible to reproduce the exact TD 
longitudinal current in a system of non-interacting electrons. In 
Ref.~\onlinecite{StefanucciAlmbladh:04-2} it was shown 
that if a steady-state is reached in the long-time limit then the current 
is given in terms of a Landauer-like formula with KS bias 
$V_{\alpha}+V_{\alpha,\rm xc}$, where $V_{\alpha}$ is the physical bias 
and $V_{\alpha,\rm xc}$ is the xc correction.
Then, to first order the current becomes (see also Ref. 
\onlinecite{KoentoppBurkeEvers:06})
\be
I=G_{0}(V_{L}-V_{R}+V_{L,\rm xc}-V_{R,\rm xc})\int d\w \frac{\partial f_{\beta}(\w)}{\partial \w}{\cal T}(\w).
\label{exI}
\ee
Consequently, the finite temperature conductance $G=I/(V_{R}-V_{L})$
 coincides with that of the
DFT+NEGF approach in Eq.~(\ref{conduct_zero}) only provided that 
$V_{\alpha,\rm xc}=0$. The exact expression for the xc bias is (to 
first order in $V_{\alpha}$)
\be
V_{\alpha,\rm xc}=
\lim_{i\to\infty}\sum_{r} f_{\rm 
xc}(i\alpha,r)\delta 
n_{r}
\label{xcbias}
\ee
where the sum runs over all sites $r$ of the model, $\delta 
n_{r}$ is the first-order density change in site $r$ and $f_{\rm 
xc}(r,r')$ is the zero-frequency xc kernel of TDDFT. In  most commonly used 
local approximations $f_{\rm xc}(r,r')\propto \delta_{rr'}$ and hence
$V_{\alpha,\rm xc}$ vanishes since $\delta n_{r}=0$  deep inside 
the leads; as a result, it is  often neglected altogether. 
The relevance of the dynamical xc correction has long been 
debated \cite{SchmitteckertEvers:08,dynxcapprox}.
The second important finding of this Letter is that at finite 
temperature the dynamical xc corrections are absolutely essential. 
They are not only important for 
the correct suppression of the Kondo peak but also for a quantitive 
description of the CB side peaks.

For the Anderson model we can find an explicit form of the dynamical 
xc correction to $G$ 
in terms of the xc kernel. The 
linear density change at site $r$ is 
\be
\delta n_{r}=\sum_{r'\in L}P_{rr'}V_{L}+\sum_{r'\in 
R}P_{rr'}V_{R}+P_{r0}\frac{U}{2}\delta n_{0}
\label{dnr}
\ee
where $P$ is the zero-frequency polarization. In linear response 
TDDFT $P$ can be 
calculated from the xc kernel as $P=P_{0}+P_{0}f_{\rm xc}P$, $P_{0}$ 
being the non-interacting polarization. Substituting Eq. (\ref{dnr}) 
into Eq. (\ref{xcbias}) and exploiting the symmetry of the $L$ and 
$R$ leads 
we find from Eq. (\ref{exI}) that the exact conductance 
 at any temperature reads 
\be
\frac{G}{G_{0}}=-(1+Q_{\rm xc})\int d\w \frac{\partial f_{\beta}(\w)}{\partial 
\w}{\cal T}(\w).
\ee
Here ${\cal T}(\w)$ is the KS transmission of Eq. (\ref{kstrans}) and
\be
Q_{\rm xc}=\lim_{i\to\infty}\sum_{r}\sum_{r'\neq 0}(-1)^{\epsilon_{\alpha}(r')}f_{\rm 
xc}(i\alpha,r)P_{rr'}
\ee
with $\epsilon_{\alpha}(r')=0$ if $r'\in\alpha$ and 1 otherwise. At 
$T=0$ we have $Q_{\rm xc}=0$ but for $T\gtrsim T_{\rm K}$ it must be
$Q_{\rm xc}\gtrsim -1$. The quantity $1+Q_{\rm xc}$ 
can be interpreted as a KS dielectric function as it measures the 
ratio between the KS and the Hartree screening.

In conclusion we proposed a finite-temperature DFT scheme to construct approximate 
xc functionals for correlated systems weakly connected to
leads. The resulting KS potential in the zero-temperature limit exhibits a 
discontinuity at integer number of particles, as it should. We showed 
that the discontinuity is essential to reproduce the conductance 
plateau due to the Kondo effect within the DFT+NEGF approach
\cite{BergfieldLiuBurkeStafford-remark}. For temperatures larger than 
$T_{\rm K}$, however, the exact KS conductance overestimates the exact 
conductance by an order of magnitude. We traced back the origin of 
this problem to the lack of dynamical xc corrections which we expect
to be 
relevant not only in this context but also in the description of 
finite temperature and finite bias transport experiments of weakly coupled 
molecular junctions. 

S.~K.~ acknowledges funding by the "Grupos Consolidados UPV/EHU del 
Gobierno Vasco" (IT-319-07) and the European Community's Seventh Framework 
Programme (FP7/2007-2013) under grant agreement No. 211956.

\end{document}